\documentclass[aps,twocolumn,reprint,nofootinbib,floatfix]{revtex4-1}
\usepackage{microtype} 

\usepackage{amsmath,amstext, amssymb, amsthm, amsfonts, slashed}
\usepackage{physics}
\usepackage{mathtools}
\usepackage[version=4]{mhchem}
\usepackage{graphicx}
\usepackage{booktabs}
\usepackage{dcolumn}% Align table columns on decimal point
\usepackage{bm}% bold math
\usepackage{bbm}
\usepackage{dsfont}
\usepackage[usenames,dvipsnames]{xcolor}
\usepackage[normalem]{ulem}
\usepackage{footmisc}

\usepackage{url}
\usepackage[colorlinks=true,urlcolor=blue,linkcolor=blue,citecolor=blue,hypertexnames]{hyperref}
\usepackage[nameinlink]{cleveref}
\usepackage{braket}
\usepackage{simplewick}
\usepackage{float}

\hypersetup{
	colorlinks,
	linkcolor={blue!75!black},
	citecolor={blue!75!black},
	urlcolor={blue!75!black}
}

\newcommand{\commentmute}[1]{} %%mute

\graphicspath{{./Figures/}}
\begin{document}

%%%%%%%%%%%%%%%%%%%%%%%%%%%%%%%%%%%%%%%%%%%
% opening
%%%%%%%%%%%%%%%%%%%%%%%%%%%%%%%%%%%%%%%%%%%
\title{Unveiling New Phases of the Standard Model Higgs Potential}

\author{Florian~Goertz}
\affiliation{Max-Planck-Institut f\"ur Kernphysik P.O. Box 103980, D 69029, Heidelberg, Germany}
\author{\'Alvaro~Pastor-Guti\'errez}
\affiliation{Max-Planck-Institut f\"ur Kernphysik P.O. Box 103980, D 69029, Heidelberg, Germany}

\begin{abstract}
We present evidence for new phases of the Standard Model Higgs potential. We study the Standard Model physical trajectory accounting for the Higgs curvature mass with the mass-dependent functional renormalisation group. New unstable and non-trivially stable phases are found at energies above the Planck scale and below the Abelian Landau pole. While the first aggravates the well-known metastable phase and threatens the viability of the Standard Model extrapolated to arbitrary scales, the latter can provide a well-defined ultraviolet completion. We investigate the phase diagram as a function of the top quark pole mass and study the effect of new physics through a scalar singlet portal coupling. The new non-trivial phase appears below the Planck scale in extensions of the Standard Model seeking stable trajectories. These findings have a significant impact on existing model building.
\end{abstract}

\maketitle

%%%%%%%%%%%%%%%%%%%%%%%%%%%%%%%%%%

%
\section{Introduction}
The Standard Model (SM) currently stands as the most successful theory in describing high-energy particle physics. Therefore, it might be considered the ultimate fundamental theory of nature describing all physical phenomena at arbitrary energy scales. For this purpose, the dependence of the SM parameters on the energy scale needs to be resolved by accounting for the quantum corrections of the full theory. With the discovery of the Higgs boson~\cite{Aad:2012tfa,CMS:2012qbp}, all free parameters of the SM were measured allowing for the high-energy behaviour of the couplings to be reconstructed to good accuracy.

However, there exist shreds of evidence of various natures towards the incompleteness of the SM. From the phenomenological side, dark matter and the excess of matter over anti-matter in the universe do not find an explanation within the SM. From renormalization-group (RG) studies of the SM, the quartic Higgs coupling is known to become negative at large energy scales \cite{Lindner:1985uk,Buttazzo:2013uya,Degrassi:2012ry,EliasMiro:2011aa,Devoto:2022qen}, leading to the possibility of the current electroweak (EW) vacuum tunnelling to a different global minimum. Additionally, the SM only includes three of the four fundamental forces known to exist. Quantum gravity effects are not englobed and expected to be relevant at the Planck scale $k_{\rm Planck}=10^{19}\,{\rm GeV}$. Last, the hypercharge coupling increases with energy and finally diverges at the Landau pole scale. This signals that the SM in its minimal formulation cannot furnish a fundamental theory valid at arbitrary energies. 

In this work, we revisit the SM high-energy trajectory employing the functional Renormalisation Group (fRG)~\cite{Wetterich:1992yh,Ellwanger:1993mw, Morris:1993qb}. This RG method allows studying the scale dependence of the Higgs curvature mass which provides information on the phase structure of the Higgs potential in RG scale and on phenomena such as spontaneous symmetry breaking (SSB). This conceptually different treatment of the mass parameter with respect to standard perturbative RG methods brings insight into new high-energy phases of the Higgs potential.

\section{Functional renormalisation group}
We employ the fRG to derive the scale dependence of the fundamental couplings in the SM - for a review see~\cite{Dupuis:2020fhh}. This non-perturbative Wilsonian RG \cite{Wilson:1970ag,Wilson:1974mb} based on the effective action formalism, consists of introducing a mass-like regulator term $R_k$ at the level of the classical action which allows for a progressive integration of momentum shells. The fRG scale $k$ acts as an infrared (IR) cutoff and can be understood as an average physical momentum or temperature, see eg.~\cite{Helmboldt:2014iya,Fu:2019hdw}. The evolution of the effective average action along the RG flow is described by the \textit{flow}, or \textit{Wetterich equation} \cite{Wetterich:1992yh}. This equation exhibits very useful properties, one of them being particularly beneficial in the current investigation: It is a mass-dependent renormalisation scheme that grants the inclusion of threshold effects. This permits us to treat the curvature mass (curvature of the potential at small field values) as a flowing parameter. 

\section{Flowing Standard Model Higgs potential.}
For the sake of simplicity we consider a polynomial Higgs effective potential, whose dimensionless and renormalised version reads
\begin{align}\label{eq:ueff}
	u(\bar\rho ) =V_{\rm eff}(\rho)\,/ \,k^4= \bar \mu^2 \, \bar \rho +\bar \lambda \, \bar \rho^2 \,,
\end{align}
where 
\begin{align}\label{eq:barrho}
	\bar\rho &= Z_\Phi \frac{ \tr \Phi^\dagger \Phi}{k^2}\,, 
	&\text{with}
    &
	&\Phi= \frac{1}{\sqrt{2}}\begin{pmatrix} \mathcal{G}_1 +i \mathcal{G}_2 \\   H  +i \mathcal{G}_3  \end{pmatrix} \,.
\end{align}
Here $H,\, {\cal G}_{1,2,3}$ denote the Higgs and Goldstone real scalar fields, $Z_\Phi$ the wave function renormalisation of the scalar doublet and $\bar \mu^2$ and $\bar \lambda$ the dimensionless and renormalised versions of the Higgs curvature mass and the quartic self-coupling. The minimum of the potential is defined as
\begin{align}\label{eq:minimumderivation}
 \left.	\partial_{\bar \rho}\, u (\bar\rho)\right|_{\bar\rho_0}=0=  \bar \mu^2  +  2\, \bar \lambda \,  \bar  \rho_0
\end{align}
with $\bar\rho_0= \bar v^2/2 =-\bar\mu^2/ (2 \, \bar \lambda )\geq 0$ and $\bar v$ the Higgs vacuum expectation value.

The scale dependence of all scalar sector parameters can be extracted from the flow of the effective potential by performing field derivatives. The flow of the dimensionless Higgs curvature mass is obtained by performing one $\bar\rho$ derivative and evaluating at the minimum of the potential,
\begin{align}\label{eq:flowmu}
\partial_t \,\bar \mu^2&=\partial_t\left(\partial_{\bar \rho} u(\bar \rho) \right)=\left(-2+ \eta_\Phi  \right) \bar \mu^2 + \partial_{\bar \rho} \,\,\overline{\text{Flow}}\left[	V_{\rm{eff}}\right]
\end{align}
where $\partial_t=k\,\partial_k$, $\,\overline{\text{Flow}}\left[V_{\rm{eff}}\right]=\left(\partial_t|_{\rho}V_{\rm{eff}}(\rho)\right)/k^4$ is the diagramatic flow of the effective potential carrying the quadratic divergence and $\eta_\Phi=-\partial_t~Z_\Phi/Z_\Phi$ is the anomalous dimension of the scalar doublet. In \eqref{eq:flowmu} we show only the flow in the symmetric phase where $\bar\rho_0=0$. For a detailed derivation, we refer to the Supplemental Material. 

While the two-loop universal scale dependence of the marginal coupling $\bar\lambda$ is well known from perturbative computations, the renormalisation of relevant parameters such as the Higgs mass is subject to scheme dependence at one loop. Hence, different schemes decrypt different features along the RG flow and connect differently to physical quantities (eg. pole mass). Employing the mass-dependent fRG allows to study the RG-flow of the curvature mass, tracking modifications in the shape of the effective potential and thereby giving reliable access to the phase structure.

\begin{figure*}[t!]
	\centering
 	\includegraphics[width=\textwidth]{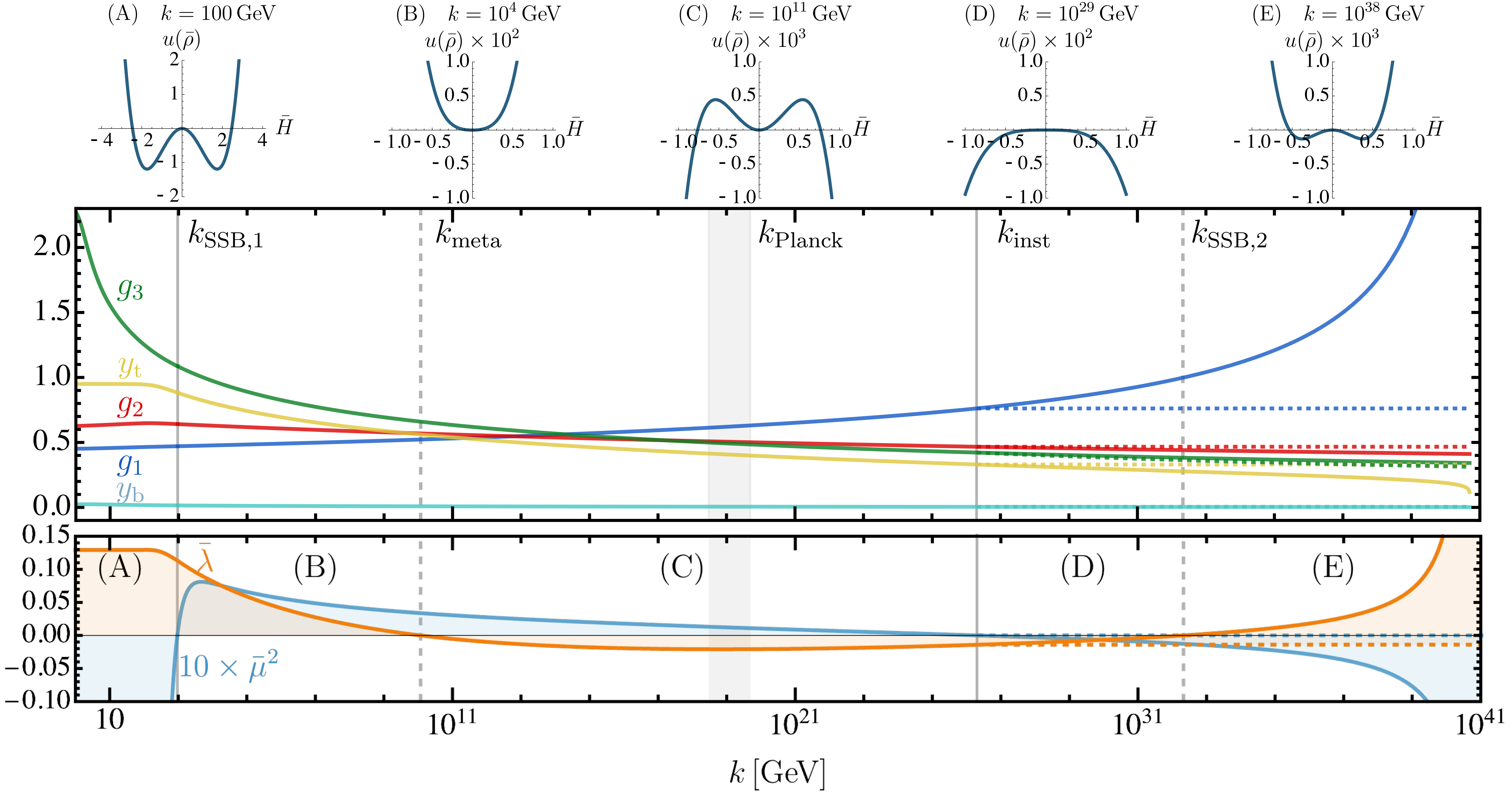}
	\caption{In the central panel, the physical RG trajectories for the renormalised SM gauge, top and bottom Yukawa couplings are displayed. These span from the Abelian Landau pole scale ($k_{\rm LP}\approx 10^{40}$\,GeV) down to below the dynamical chiral symmetry breaking scale ($k_{\chi{\rm SB}}\approx 0.1$\,GeV), over the Planck ($k_{\rm Planck}\approx 10^{19}$\,GeV) and EW ($k_{\rm SSB, 1}\approx 10^{3}$\,GeV) scales. In the bottom panel, the RG flow of the dimensionless Higgs curvature mass and the quartic coupling are shown. These two parameters determine the shape of the effective potential and hence the phase of the Higgs field. Five different regions are separated by plain ($\bar \mu^2=0$) and dashed ($\bar \lambda=0$) vertical lines. In the top panel, we plot the phases of the dimensionless potential as a function of the dimensionless Higgs field $\bar H= H/k$. The dashed trajectories above $k_{\rm inst}$ correspond to the decoupling scenario discussed in the text.}
	\label{Fig:SMtrajectory}
\end{figure*}

\section{Standard Model trajectory and transplanckian Higgs phases}

In \cite{Pastor-GutierrezSciPostPhys.15.3.105}, the flows of all SM parameters along the effects of asymptotically safe quantum gravity~\cite{Weinberg:1980gg, Reuter:1996cp,Souma:1999at} were computed using the fRG. In~\cite{Borchardt:2016xju,Sondenheimer:2017jin,Gies:2013fua,Gies:2013pma,Gies:2016kkk,Gies:2017ajd,Sondenheimer:2017jin,Gies:2019nij,Gies:2018vwk,Gies:2015lia}, the phase and fixed point structure of SM-like theories had previously been studied with functional methods. Here, we analyse the SM and neighbouring trajectories without the consideration of quantum gravity effects. The physical trajectory reproducing the experimentally measured SM values is presented in \Cref{Fig:SMtrajectory}. The flows span from the Landau pole scale $k\approx 10^{41}$\,GeV, over the Planck and the EW scales, down to the deep IR, below the QCD scale. For details on the global truncation, scale setting procedure and systematics estimate we refer to the Appendices of \cite{Pastor-GutierrezSciPostPhys.15.3.105}. 

In the lower panel of \Cref{Fig:SMtrajectory} we show the Higgs potential parameters. The curvature mass is negative below $k_{\rm SSB, 1}\approx 1$\,TeV meaning that the potential has a non-trivial minimum (see (A) in the top panel). As the fRG flows are sensitive to threshold effects, the progressive decoupling of degrees of freedom causes the freeze out of the couplings below the respective mass scales.

At scales above $k_{\rm SSB, 1}$, $\bar\mu^2>0$ and $\bar\lambda>0$ lead to a stable Higgs potential (B) with a single minimum at $H=0$. Here, all mass terms, with the exception of the curvature mass, vanish. In the absence of threshold effects, all marginal coupling flows recover their universal analytic structure. This is appreciable in the scaling profile above $k_{\rm SSB,\,1}$ which agrees with perturbative computations~\cite{Buttazzo:2013uya,Hiller:2022rla}. An important well-known scale is where the scalar quartic self-coupling evolves to negative values~\cite{Lindner:1985uk,Casas:1994qy,Isidori:2001bm,EliasMiro:2011aa,Devoto:2022qen,Degrassi:2012ry,Bezrukov:2012sa,Bednyakov:2015sca,Buttazzo:2013uya}. Above $k_{\rm meta}\approx 1.2 \cdot10^{10}$\,GeV the Higgs field shows a local minimum at $H=0$ but is unbounded from below (potential (C)).

In the current work we are interested in the features of the SM as a pure matter Quantum Field Theory. For the sake of studying the interplay of the different sectors, we disucuss the couplings profile above $k_{\rm Planck}$. This assumption is compatible with the consideration of a weak gravity scenario in which the quantum gravitational effects do not lead to a qualitative change in the matter sector. Furthermore, in the upcoming sections we will show how when considering well-motivated new physics extensions, the effects discussed here appear below the Planck scale. Therefore, for the first time we find that $\bar\mu^2$ turns negative at $k_{\rm inst}\approx 2\cdot 10^{26}$\,GeV meaning that the curvature of the Higgs potential at small field values is negative which, together with a negative quartic coupling, leads to an unstable potential ((D) in \Cref{Fig:SMtrajectory}). 

One may speculate that as this phase is reached the Higgs field evolves towards infinite (or very large) values causing the decoupling of all degrees of freedom besides the photon and the gluons. Consequently and given the Abelian nature of $g_1$, all couplings except $g_3$ reach a non-trivial fixed point as $\bar \mu^2$ reaches $(\bar \mu^2)^*=0$. This provides an ultraviolet (UV) complete picture from which the SM emerges out of an asymptotically free SU(3) Yang-Mills theory. This is depicted in \Cref{Fig:SMtrajectory} above $k_{\rm inst}$ by dashed lines. 

This unstable potential phase comes along with severe consequences that question the validity of the SM in this minimal approximation. First, the ground state of the Higgs field is not well defined, hence the theory. Second, as $\bar \lambda$ does not return to positive values, this scenario leads to a ruled-out EW vacuum decay probability~\cite{Kobzarev:1974cp,Coleman:1977py,Callan:1977pt,Isidori:2001bm,Devoto:2022qen,Buttazzo:2013uya,Branchina:2013jra,Branchina:2014rva,Markkanen:2018pdo}. Last, recalling that the potentials displayed in the top panel of \Cref{Fig:SMtrajectory} can be conceptually understood as the thermal phase evolution, the transition from an unstable to a metastable potential cannot be realised without a $\bar \lambda>0$. Altogether, these findings point out that the phenomenological and formal validity of the SM could be at stake at scales way below the Landau pole. 

For illustrative purposes, we may assume that the Higgs field remains at $H=0$ along the unstable regime, so that the flows can be continued towards higher energies. This phase spans over 6 orders of magnitude until the quartic coupling returns to positive values at $k_{\rm SSB, 2}\approx 2\cdot 10^{32}$\,GeV developing a stable shape with a non-trivial minimum (potential (E)). This non-trivial potential gives rise to gauge boson and fermion masses in a controlled and well-defined manner and as will be discussed later on, could provide a cure for the Landau pole. 

It is important to stress that in \eqref{eq:ueff} we have considered a polynomial Higgs potential only containing renormalisable operators. We note that an unstable shape of the potential at large field values could be resolved by including higher order operators such as $\bar\rho^3\,,\bar\rho^4\,,\ldots$ in the expansion~\cite{Fu:2019hdw,Pastor-GutierrezSciPostPhys.15.3.105,Borchardt:2016xju,Gies:2018vwk,Gies:2016kkk}. In the absence of a UV completion, the respective couplings are free, leading to a highly complex analysis beyond the scope of this work and which will be addressed in an independent work. The addition of these is however not expected to change the behaviour of the well-converged small field curvature, given that this is dominantly determined by gauge and Yukawa effects, which is at the centre of our analysis. Beyond that, the potential relevance of higher terms highlights that the decoupling (dashed) and the illustrative $H=0$ (plain) scenarios can be seen as two limiting cases of a well-behaved and stabilised setup, with large and small background value, respectively. Crucially, the presence of a new UV phase due to the zero in the quadratic term is a robust and new prediction (independent of the presence of higher terms). As mentioned, further below we will show how the same physics, embedded in a slightly modified new physics scenario, leads to the emergence of the new stable phase below the Planck scale, without intermediate instabilities and thereby avoiding such subtleties at large field values.

\begin{figure}[t]
	\centering
	\includegraphics[width=	1. \columnwidth ]{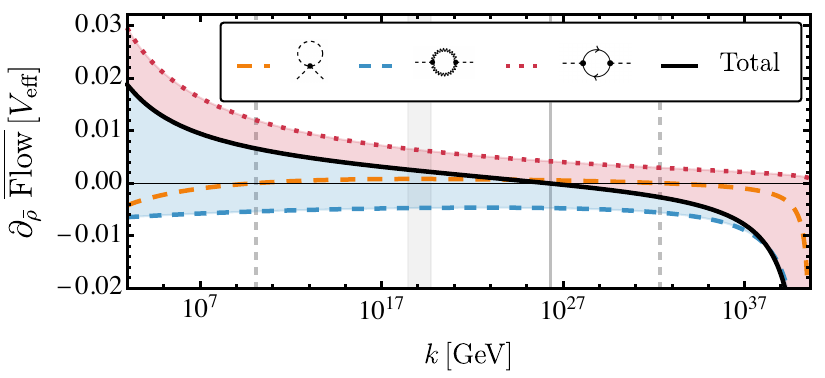}
	\caption{Total (solid black line) and partial (dashed lines) contributions of the scalar (orange), gauge (blue) and fermion (red) loops to the first derivative of the flow of the effective potential relevant for the flow of the curvature mass. Vertical lines delimit the different phases as in \Cref{Fig:SMtrajectory}.}
	\label{Fig:contributionsflowmu}
\end{figure}
\section{Ultraviolet zero-crossings of the curvature mass}\label{sec:UVzero crossing}
To understand what triggers a $\bar \mu^2$ zero-crossing which leads to the emergence of new phases, we analyse the contributions in the flow \eqref{eq:flowmu}. These can be separated in two: the proportional to the curvature mass and the independent. The former is driven by the canonical scaling and the anomalous dimension of the Higgs field. Importantly, these corrections drive the curvature mass asymptotically towards zero but do not induce a zero-crossing. For this, a negative contribution independent of $\bar \mu^2$ must be present in \eqref{eq:flowmu}. This, in fact, appears in the quadratic divergence of the Higgs two-point function flow which in \eqref{eq:flowmu} is shown as the $\bar\rho$-derivative of the effective potential flow. The leading contributions in the Landau gauge read 
\begin{align} \label{eq:contributionsflowmu}
\hspace{-0.3 cm}\partial_{\bar \rho}\,\overline{\text{Flow}}\left[	V_{\rm{eff}}\right]\supset \frac{3\, y^2_t}{8 \, \pi^2}  -\frac{9\left( g^2_1 +5 \,g^2_2\right)}{320 \, \pi^2}- \frac{3\bar \lambda}{8\,\pi^2(1+ \bar \mu^2)^2}  \,.
\end{align}

Each term in \eqref{eq:contributionsflowmu} and its total are depicted in \Cref{Fig:contributionsflowmu}. 
The Yukawa-mediated diagram contributes positively to supporting the symmetric phase while the gauge contributions drive towards negative values to induce SSB. Given that the $\bar\mu^2$-dependent contributions already drive $\bar \mu^2$ close to zero, at $k_{\rm inst}\approx 10^{26}$\,GeV the potential becomes unstable as the total $\bar \mu^2$-independent contribution in \eqref{eq:contributionsflowmu} turns negative. We can determine the $\bar \mu^2$-zero crossing scale $k_{\rm cross}=\{k_{\rm inst}\,|\,k_{\rm SSB,\,2}\}$ analytically via the condition
\begin{align}\label{eq:relationkSSB2}
	\left.y^2_{\rm t}(k)\right|_{k_{\rm cross}} \simeq \left. \left[ \frac{3}{40}\big[g_1^2(k) + 5\, g_2^2(k)\big]+ \bar \lambda(k) \right]\right|_{k_{\rm cross}}\,.
\end{align}
As the total gauge contribution is approximately constant over all scales and the top Yukawa coupling weakens, a $\bar\mu^2$-zero crossing towards negative values is hardly avoidable. Importantly, this equality is generally satisfied at scales where all couplings are perturbative and hence the one-loop structure suffices to determine $k_{\rm cross}$. For this reason, further zero crossings to positive $\bar \mu^2$ are unlikely to occur. 

It is essential to stress that given the irrelevance of the threshold effects in \eqref{eq:contributionsflowmu} and \eqref{eq:relationkSSB2} the results here discussed hold independently of the regulator choice and are common to any Callan-Symanzik-like scheme \cite{Symanzik:1970rt,PhysRevD.2.1541} where the regularization procedure suppresses modes in a mass-like fashion.

\begin{figure}[t]
	\centering
	\includegraphics[width=\columnwidth]{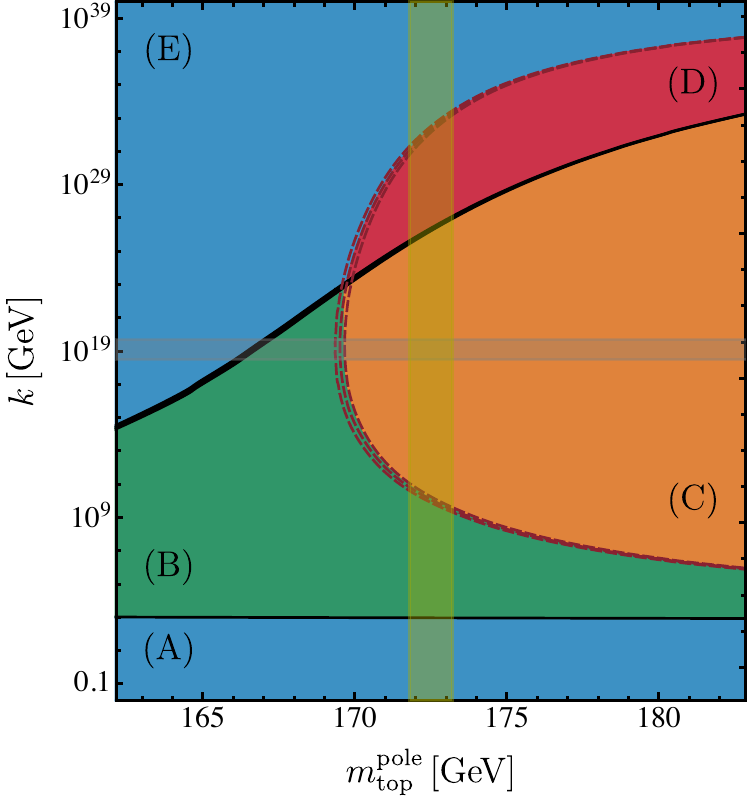}
	\caption{Phase diagram of the SM Higgs potential for different top quark pole masses. The plain black (dashed red) lines show the scales at which the curvature mass (quartic coupling) vanishes for trajectories reproducing the measured Higgs pole mass and $\pm 3\%$ variations. The blue (green) region depicts the regime in which the Higgs potential is stable and has a non-vanishing (vanishing) minimum. The red (orange) region denotes the unstable (metastable) regime. The Planck scale is marked as a grey-shaded horizontal region and the top pole mass experimental uncertainty \cite{ParticleDataGroup:2020ssz} as a vertical yellow-shaded band.}
	\label{Fig:SMphasediagram}
\end{figure}
\section{New phases and need for new physics}\label{sec:NewPhasesfornewphysics}
We have discussed how a $\bar\mu^2$-zero crossing is inevitable in the pure SM framework and therefore a negative quartic coupling not only leads to a metastable but also to a potentially severely problematic unstable potential. Therefore, addressing the meta/instability of the Higgs potential becomes a highly relevant task. The main negative contribution to the flow of the effective potential is sourced in the large top Yukawa coupling measured in the IR. As the fermionic loops screen $\bar\lambda$, decreasing $y_t$ leads to a stabilisation of the potential. Given that the gauge coupling flows are insensitive to these variations, such an approach to a stabilisation of the Higgs potential causes a $\bar \mu^2$-zero-crossing at much lower scales. This investigation is summarised in \Cref{Fig:SMphasediagram}. Although the top quark pole mass measurement \cite{ParticleDataGroup:2020ssz} shows the largest uncertainty in all SM parameter measurements, the physical trajectory leads confidently to an unstable regime. For a $m^{\rm pole }_{\rm top}=169$\,GeV ($2\%$ smaller than the measured value) the quartic coupling is always positive and the $k_{\rm SSB, 2}$ scale has decreased four orders of magnitude. Moreover, for a $m^{\rm pole }_{\rm top}=166$\,GeV ($4\%$ smaller than the measured value) $k_{\rm SSB, 2}<k_{\rm Planck}$ and of order of inflationary scales.

\begin{figure}[t]
	\centering
	\includegraphics[width=	\columnwidth]{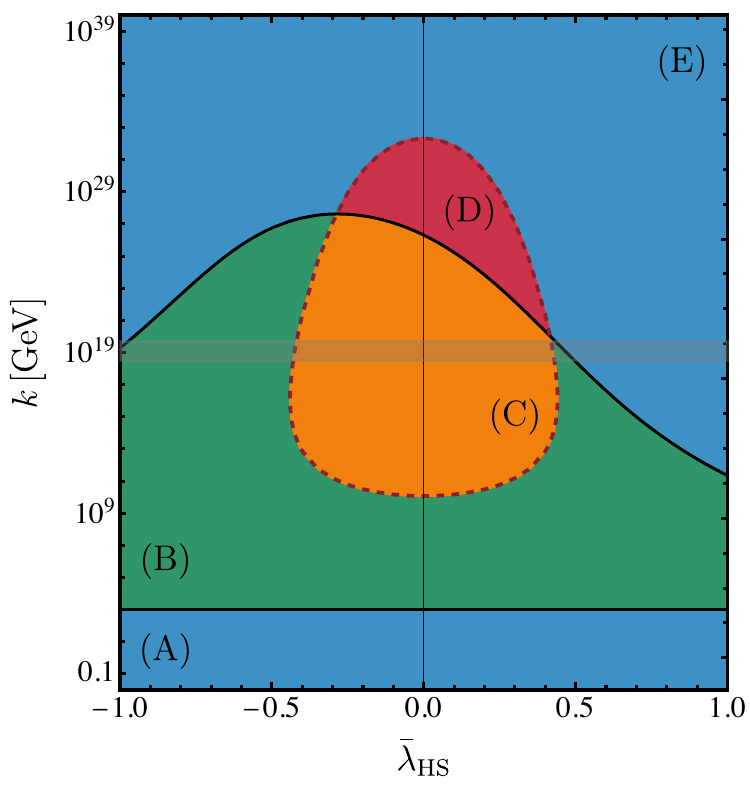}
	\caption{Phase diagram of the SM Higgs potential with an additional singlet scalar field as a function of the portal coupling \eqref{eq:portal}. Same color coding and labeling as in \Cref{Fig:SMphasediagram}.}
	\label{Fig:SMphasediagramportal}
\end{figure}

New physics beyond the SM can provide an explanation to standing open problems and additionally stabilise $\bar \lambda$~\cite{Hiller:2020fbu,Hiller:2022rla,Litim:2015iea,Held:2018cxd,Branchina:2013jra,Branchina:2014rva,Bally:2022naz,Bally:2023lji}. As an example, we may consider a minimal extension with a real singlet scalar $S$, coupled only to the Higgs field via a portal coupling \cite{Arcadi:2019lka,Espinosa:2011ax,Elias-Miro:2012eoi} 
\begin{align}\label{eq:portal}
	\Delta u (\bar\rho,\bar\rho_{\rm S})= \bar \lambda_{\rm HS} \,\bar \rho \, \bar\rho_{\rm S} \,,
\end{align}
where $\bar \rho_{\rm S}= Z_S \,S^2/2$. Being agnostic about the singlet field potential, we set different values of the portal coupling above the threshold of the singlet mass scale assumed to be of the EW scale. In \Cref{Fig:SMphasediagramportal}, we summarise the effect of this minimal extension on the phase structure of the Higgs potential. For $|\bar \lambda_{\rm HS}|\gtrsim 0.4$ the positivity of $\bar \lambda$ is ensured at all energy scales. As the corrections to the Higgs 2-point function are linear in $\bar\lambda_{\rm HS}$, there is no symmetric dependence in $k_{\rm cross}$. For positive couplings, the new UV broken phase emerges at $k_{\rm SSB, 2} < k_{\rm Planck}$ and for negative portal couplings this is pushed to higher energy scales. Moreover, for $-0.4\lesssim \bar \lambda_{\rm HS}\lesssim -0.3$ trajectories with metastable but no unstable regions can be found.

\section{Towards a fundamental Standard Model}
We have shown that the SM trajectory leads to an unstable potential which could either mean that the theory is ill-defined and hence physically unviable or further minima exist but cannot be resolved in the current lowest-order polynomial approximation. In this last case and in stable SM trajectories as the ones discussed, the only impediment against formulating a fundamental theory valid at arbitrarily large energy scales is the presence of a Landau pole in the Abelian coupling. This divergence is core-rooted in the quantum corrections to $g_1$ mediated by massless fermions and bosons. This strong regime will now be approached in the presence of a non-trivial $\bar v$ and where, as in the IR, the relevant degrees of freedom are properly described by a massless photon and massive $Z^0$, $W^\pm$ and fermion fields. 

The UV non-trivially stable regime discussed in this work provides a natural way to cure the Landau pole and formulate the SM as a fundamental theory. To avoid this divergence, the quantum corrections to the electric coupling must vanish or become negative, achieving asymptotic safety or freedom. The first case is realised with the decoupling of all electrically charged fermionic fields as the quantum fluctuations of momenta $k$ being integrated out are smaller than the euclidean masses of the off-shell particles propagating, $m_{\psi_i} > k$. As the fermion masses are generated through the Higgs mechanism, this constraint translates into the UV fixed point condition 
\begin{align}\label{eq:decouplingFP}
 \bar v^* y^*_i /\sqrt{2}> 1 \,.
\end{align}
\begin{figure}[t]
	\centering
	\includegraphics[width=	1. \columnwidth ]{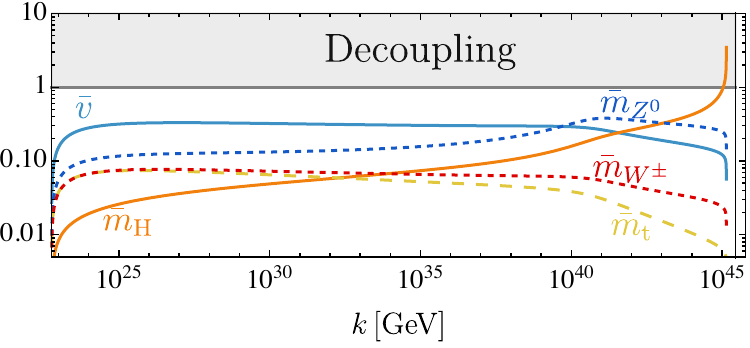}
	\caption{Scale dependence of $\bar v$ (plain blue) and Higgs (plain orange), $Z^0$ (blue dotted), $W^\pm$ (red dotted) and top quark (yellow dashed) euclidean masses above $k_{\rm cross}$ in a stable SM-like trajectory with a lighter top quark pole mass of $169.4$\,GeV. }
	\label{Fig:UVbrokenphaseflows}
\end{figure}

In \Cref{Fig:UVbrokenphaseflows} we display as an example, the scale dependence of the dimensionless masses and the flowing minimum for a stable SM-like trajectory where $m^{\rm pole}_{\rm top}=169.4$\,GeV. 
Although the potential develops a $\bar v\approx 0.3$, this trajectory does not suffice to trigger the decoupling of any of the fermionic fields as all Yukawa couplings also remain small. However, the presence of masses over many orders of magnitude leaves an imprint on the couplings which is noticeable in the Landau pole scale being delayed by five orders of magnitude from the scenario with $\bar v=0$. Nevertheless, for a conclusive resolution of the evolution along the strong Abelian regime and its impact on the other sectors, an improved truncation is necessary. Besides, \eqref{eq:decouplingFP} is trivially satisfied if $\bar v\to \infty$ as shown in \Cref{Fig:SMtrajectory}. In the case of further hidden minima unresolvable in the current polynomial approximation, the question remains open. 

\section{Conclusions}
In this work, we have studied the SM trajectory at all accessible energy scales using the fRG. The treatment of the Higgs mass parameter with this Wilsonian RG allows the evolution of the full effective Higgs potential along the RG flow and to trace occurrences as SSB. 

In the extrapolation of the SM trajectory from measured values, the Higgs potential undergoes previously unknown phases: an unstable and a stable with a non-trivial minimum. The scale where the curvature mass shows a zero crossing is triggered by the gauge corrections overtaking the top Yukawa's and can be determined by a simple equality \eqref{eq:contributionsflowmu}. This crossing occurs where $\bar \lambda<0$ and leads to an unstable potential. While this finding seems to invalidate the SM trajectory at higher energy scales and signals the need for new physics, the probability of further minima unresolvable in the current polynomial approximation of the Higgs potential exists. 

We have studied the SM phase diagram for different top quark pole masses and considering new physics in the form of a Higgs-scalar portal coupling. The new UV non-trivially stable phase appears below the Planck scale in scenarios seeking stable trajectories. This phase provides a potential UV-complete formulation to the SM and similar theories via the decoupling of fermions. These findings have a great impact on existing approaches to new physics and suppose a new look at the high-energy structure of the SM.

\vspace{2cm}
{\textbf{Note added:}} A similar phase transition has been reported in~\cite{Hawashin:2024dpp}, where theories involving two types of scalar fields with ${\rm O}(N) \times \mathbb{Z}_2$ symmetries in $2+1$ dimensions, including finite-temperature effects, have been studied. It is shown that the transition from the symmetric to the broken phase, analogous to the one discussed in \Cref{sec:UVzero crossing}, is driven by the interplay between scalar contributions and power-law corrections to the flow of the curvature mass. This result supports the findings of the present work.

\begin{acknowledgments}
We would like to thank Andrei Angelescu, Andreas Bally, Holger Gies, Jan M. Pawlowski, Manuel Reichert, Richard Schmieden, Christof Wetterich, Masatoshi Yamada and Luca Zambelli for fruitful discussions and comments on the manuscript. 
\end{acknowledgments}

\appendix

\begin{widetext}
\section*{Appendix}

In this Appendix, we provide further details on the flows of the SM Higgs potential parameters. These can be obtained by performing field derivatives of the flow of the effective potential and evaluating at its minimum. This reads
\begin{align}\label{eq:flownderu}
	\left.\partial_t \left( \partial^n_{\bar \rho}\, u (\bar \rho) \right)\right|_{\bar \rho_0} &= \partial^n_{\bar \rho} \left[\left.\partial_t \right|_{\rho}u(\bar \rho)- \left.\partial_t \right|_{\rho} \bar \rho \cdot \partial_{\bar \rho} u(\bar \rho) \right]_{\bar \rho_0}+ \partial_t \bar \rho_0\cdot \left.\partial^{n+1}_{\bar \rho} u(\bar \rho)\right|_{\bar \rho_0}\,,
\end{align}
where we have rewritten the scale derivative for a fixed $\bar\rho$. Now, employing the definition of the dimensionless diagrammatic flow of the potential 
\begin{align} \label{eq:flow-Higgs-pot}
	\overline{\text{Flow}}\left[	V_{\rm{eff}}\right]  =  \frac{\partial_t|_{\rho}	V_{\textrm{eff}}(\rho)}{k^4}=	\left.\partial_{t}\right|_{\rho}u(\bar\rho)+ 4 \, u (\bar\rho) \, 
\end{align}
and 
\begin{align}
	\left. \partial_{t}\right|_{\rho}\bar \rho= -\left(2+\eta_\Phi\right)\bar \rho \,,
\end{align}
we reach the general expression,
\begin{align}\label{eq:flownderufinal}
	\left.\partial_t \left( \partial^n_{\bar \rho}\, u (\bar \rho) \right) \right|_{\bar \rho_0}&= \partial^n_{\bar \rho} \left[\overline{\text{Flow}}\left[	V_{\rm{eff}}\right] -4 \, u(\bar \rho) \right]_{\bar \rho_0}+ (2+\eta_\Phi)\,\partial^n_{\bar \rho} \left[\bar \rho\, \partial_{\bar \rho} u(\bar \rho) \right]_{\bar \rho_0}+ \partial_t \bar \rho_0\cdot \left.\partial^{n+1}_{\bar \rho} u(\bar \rho)\right|_{\bar \rho_0} \,.
\end{align}
For $n=1$ and evaluating at $\bar \rho_0=0$, we obtain the flow of the curvature mass in the symmetric phase shown in \eqref{eq:flowmu}. Performing an additional $\bar \rho$ derivative we obtain the flow of the quartic coupling,
\begin{align}\label{eq:flowlambda}
	\partial_t \bar \lambda = \partial_t(\partial^2_{\bar \rho}\, u(\bar \rho))=2\,\eta_\Phi   \bar \lambda + \frac{1}{2}\partial^2_{\bar \rho}\,\,	\overline{\text{Flow}}\left[	V_{\rm{eff}}\right] \,.
\end{align}
Note that the flows are evaluated at the minimum of the potential $\bar \rho_0 =\bar v^2/2$ and hence in the broken phase additional terms as the last in \eqref{eq:flownderufinal} do not vanish, leading to non-trivial contributions. These have been taken into account in the broken phase results depicted in \Cref{Fig:SMtrajectory,Fig:UVbrokenphaseflows}.

The diagrammatic flow of the effective potential depicted in \Cref{Fig:flowV} can be solved analytically with a suitable choice of regulator \cite{Litim:2001up}. In the Landau gauge and dropping the gluonic contributions, the dimensionless flow of the potential reads
\begin{align} \label{eq:flow-Higgs-pot-solved}
	\overline{\text{Flow}}\left[	V_{\rm{eff}}\right]=& \frac{1}{16\pi^2}\Bigg[\,\frac{3-\frac{\eta_{{\cal G}^\pm}}{3}-\frac{\eta_{{\cal G}^0}}{6}}{ 2\left(1+\bar\mu ^2+2 \bar\lambda  \bar\rho\right)}+\frac{1-\frac{\eta_H}{6}}{2 \left(1+\bar\mu ^2+6 \bar\lambda  \bar\rho\right)}+\frac{ \left(4+ \bar m^2_Z[\,\bar \rho\,]\right)\left(1-\frac{\eta_Z}{6}\right)}{2 \left(1+ \bar m^2_Z[\,\bar \rho\,]\right)}+\frac{\left(4+ \bar m^2_W[\,\bar \rho\,]\right) \left(1-\frac{\eta_{W^\pm}}{6}\right)}{  \left(1+\bar m^2_W[\,\bar \rho\,] \right)}\notag\\[1ex]
    &\hspace{8cm}-\sum_{q}\,\frac{6 \left(1-\frac{\eta_{q}}{5}\right)}{ \left(1+\bar m^2_q[\,\bar \rho\,]\right)}-\sum_{l}\,\frac{2 \left(1-\frac{\eta_{l}}{5}\right)}{ \left(1+ \bar m^2_l[\,\bar \rho\,]\right)}\Bigg]\,.
\end{align}

Here, $\bar m_Z^2[\,\bar \rho\,]=\frac{ 3  g_1^2 +5 g_2^2 }{10}\bar\rho$,  $ \bar m_W^2[\,\bar \rho\,]=\frac{g_2^2}{2} \bar\rho$ and $ \bar m^2_{q/l}[\,\bar \rho\,]=y^2_i \bar\rho$ are the dimensionless EW gauge boson and fermion eucledian masses generated through the Higgs mechanism. The mass-dependent nature of the fRG manifests itself in the threshold functions containing the masses, see eg. \cite{Yamada:2020bqe} for a comparison to perturbative schemes. It is apparent how the different contributions decouple as $\bar m_{i} >1$ is satisfied along the RG-flow. Furthermore, the flow of any arbitrary scalar $n$-point function can be obtained by performing field derivatives of \eqref{eq:flow-Higgs-pot-solved}.

The results of this work were computed employing $R_\xi$ gauge-fixing and are not affected by different choices of the gauge parameter, for details see \cite{Pastor-GutierrezSciPostPhys.15.3.105}. In the flows of the Higgs sector, $\bar\mu^2$ and $\bar\lambda$, the gauge dependence either is subleading. While the latter is independent at one loops given the universality of the marginal gauge coupling, $\bar\mu^2$ has a residual gauge dependence only appearing in the anomalous dimension. This is very well known and leads to the discussion of the Fröhlich-Morchio-Strocchi mechanism (see eg. \cite{Maas:2023emb}) as a potential formulation of the Higgs correlation functions in a gauge independent manner. However, we stress that the second term $ \overline{\rm Flow}\left[V_{\rm eff}\right] $ is gauge independent given that the contributions between longitudinal modes of the EW gauge bosons, EW ghosts and Goldstone modes exactly cancel. This shows how the emergence of the EW symmetry breaking and the UV phases is qualitatively unaffected by the gauge choice. 

To complete the resolution of the flows \eqref{eq:flowmu} and \eqref{eq:flowlambda}, the Higgs anomalous dimension needs to be computed. It can be derived from the flow of the respective 2-point functions, as for example in \cite{Pastor-GutierrezSciPostPhys.15.3.105}. On the other hand, the presence of the anomalous dimensions in \eqref{eq:flow-Higgs-pot-solved} makes apparent the higher loop effects accounted for by the flow equation. 
In \eqref{eq:contributionsflowmu} we have neglected the anomalous dimensions in the diagrammatic flow because, in perturbative regimes as the one here discussed, their magnitude is small and their contribution is largely suppressed. 

\begin{figure}
 \includegraphics[width=0.8 \textwidth]{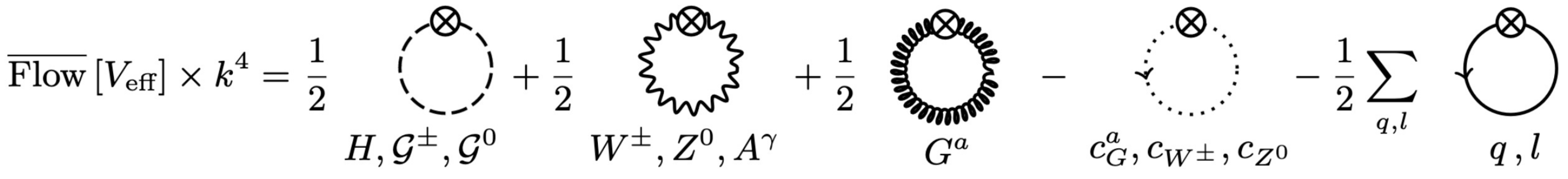}
 
	\caption{Diagrammatic flow of the SM Higgs potential. From the first to last diagram, we show the contribution of all scalar, EW gauge, gluon, ghost and fermionic loops. The lines denote full regularised propagators including all quantum fluctuations and the cut circles the insertion of the regulator scale derivative.}
	\label{Fig:flowV}
\end{figure}

We close this technical Section with a remark on the broken phase flows. In this regime, it is more convenient to define a flowing minimum of the potential,  
\begin{align}\label{eq:deru}
	\left.\partial_{\bar\rho}\, u\left(\bar\rho\right)\right|_{\bar \rho_{0}} &= 0\,,
\end{align}
where $\bar \rho_0 = \frac{ Z_\Phi v^2}{2\, k^2}=\frac{ \bar v^2}{2}=-\frac{\bar \mu^2}{2\, \bar \lambda}$. Taking a $t$-derivative, we obtain
\begin{align}\label{eq:flowrho0}
	\partial_{t}\bar \rho_{0}=-\left.\frac{ \partial_{\bar\rho} \left(\left.\partial_{t}\right|_{\rho}u(\bar\rho)-\left.\partial_t\right|_{\rho} \bar \rho \,\partial_{\bar\rho} u(\bar\rho) \right)}{\partial^2_{\bar\rho} u(\bar\rho)} \right|_{\bar \rho_0}\,.
\end{align}
This parameterization is more convenient as the $\bar v$ enters the definition of the euclidean masses generated by the Higgs mechanism and is used throughout the analysis. For example, for the Higgs field, the dimensionless euclidean mass reads $\bar m_{H} = \sqrt{2 \bar \lambda}\,  \bar v\,$ which, although still a cutoff-dependent quantity, can be linked to physical pole masses and observables (see for example the scale setting procedure in \cite{Pastor-GutierrezSciPostPhys.15.3.105}). 

\end{widetext}

%%%%%%%%%%%%%%%%%%%%%%
\bibliographystyle{apsrev4-1} 
\bibliography{references}
%%%%%%%%%%%%%%%%%%%%%%
\end{document}